\documentclass[conference,a4paper]{IEEEtran}
\usepackage{graphicx,amssymb,amsmath}
\usepackage{multicol}
\usepackage[noadjust]{cite}
\usepackage{setspace}
\usepackage{subfigure}
\usepackage{graphicx}
\usepackage{float}
\usepackage {url}
\usepackage{stfloats}
\usepackage{caption2}
\usepackage{amsthm,pifont}
\usepackage{flushend}
\usepackage{cases,subeqnarray}
\usepackage{bm,multirow,bigstrut}
\usepackage{amsmath, amsthm, amssymb}
\usepackage{textcomp}
\usepackage{latexsym,bm}
\usepackage{booktabs}
\usepackage{xcolor}
\usepackage{mathtools}
\usepackage{dsfont}
\usepackage{extarrows}
\usepackage{epsfig}
\usepackage{epstopdf}
\usepackage[noend]{algpseudocode}
\usepackage{algorithmicx,algorithm}
\theoremstyle{plain}

\theoremstyle{plain}
\newtheorem{rem}{Remark}
\newtheorem{them}{Theorem}

\usepackage{amsmath}

	\IEEEoverridecommandlockouts

\begin{document}
	\title{Optimal Targeted Advertising Strategy For Secure Wireless Edge Metaverse}
	\author{Hongyang Du, Dusit~Niyato, Jiawen~Kang, Dong~In~Kim, and Chunyan~Miao
		\thanks{H.~Du is with the School of Computer Science and Engineering, the Energy Research Institute @ NTU, Interdisciplinary Graduate Program, Nanyang Technological University, Singapore (e-mail: hongyang001@e.ntu.edu.sg).}
		\thanks{D. Niyato and C. Miao are with the School of Computer Science and Engineering, Nanyang Technological University, Singapore (e-mail: dniyato@ntu.edu.sg; ascymiao@ntu.edu.sg)}
		\thanks{Jiawen Kang is with the School of of Automation, Guangdong University of Technology, China (e-mail: kavinkang@gdut.edu.cn)}
		\thanks{D. I. Kim is with the Department of Electrical and Computer Engineering, Sungkyunkwan University, Suwon 16419, South Korea (e-mail: dikim@skku.ac.kr)}
	}
	\maketitle
	\vspace{-1cm}
	\begin{abstract}
Recently, Metaverse has attracted increasing attention from both industry and academia, because of the significant potential to integrate real and digital worlds ever more seamlessly. By combining advanced wireless communications, edge computing and virtual reality (VR) technologies into Metaverse, a multidimensional, intelligent and powerful wireless edge Metaverse is created for future human society. In this paper, we design a privacy preserving targeted advertising strategy for the wireless edge Metaverse. Specifically, a Metaverse service provider (MSP) allocates bandwidth to the VR users so that the users can access Metaverse from edge access points. To protect users' privacy, the covert communication technique is used in the downlink. Then, the MSP can offer high-quality access services to earn more profits. Motivated by the concept of ``covert", targeted advertising is used to promote the sale of bandwidth and ensure that the advertising strategy cannot be detected by competitors who may make counter-offer and by attackers who want to disrupt the services. We derive the best advertising strategy in terms of budget input, with the help of the Vidale-Wolfe model and Hamiltonian function. Furthermore, we propose a novel metric named Meta-Immersion to represent the user's experience feelings. The performance evaluation shows that the MSP can boost its revenue with an optimal targeted advertising strategy, especially compared with that without the advertising.
	\end{abstract}
	\IEEEpeerreviewmaketitle
	\section{Introduction}
The ``Metaverse", which was first proposed by Neal Stephenson in his novel Snow Crash \cite{joshua2017information}, is a hypothetical synthetic environment created by computers that are linked to the physical world. A number of companies, such as Facebook, Epic Games, and Tencent, have announced their ventures into Metaverse. Specially, the social media giant Facebook re-branded itself as ``Meta" to build the Metaverse.

For the metaverse, several promising 6G techniques, e.g., Terahertz communication and reconfigurable intelligent surfaces, can be used to provide high data rate and ultra-low latency communications. Moreover, progresses in the development of multimedia technologies, such as remote and non-panoramic rendering, make wireless virtual reality (VR) and augmented reality (AR) become the most promising technologies for accessing the Metaverse \cite{kelkkanen2020bitrate,boardieee}. From the perspective of wireless resource management, a Metaverse service provider (MSP) needs to determine the bandwidth allocated to VR users, ensuring normal-quality access to the Metaverse. The excess bandwidth can be used to provide ``high-quality access service" to users thus making more profits. Note that advertising is an effective way to promote such new businesses to entice customers to upgrade their products and services \cite{yang2020nonparametric}. 

However, MSPs may avoid their advertisement being detected by competitors and attackers. The reason is that the advertisement may contain the MSP's marketing details, target users' local information or special characters \cite{toubiana2010adnostic}. Inspired by the concept of ``covertness" in covert communication \cite{zheng2019multi}, the advertising can be hidden in noisy information \cite{anand2009targeted}, i.e., textual and interpersonal meta-discourse, to avoid potential attacks. The transmission of information can only be received by intended users. Thus, the MSP can customize and optimize the actual advertising effort and the noisy information effort to make the detection error probability (DEP) of competitors and attackers close to one. Such a strategy is called targeted advertising \cite{toubiana2010adnostic}, which can be used in many real-life cases. For example, Horizon Workrooms, a Facebook's Metaverse software, can let people have conferences in the same virtual room. Because we do not want the conference information to be compromised as well as the competitors to counter-offer similar services, the advertising of high-quality access should not be detected by people other than participants.


To address the aforementioned problems, we propose a targeted advertising management framework of high-quality access, and investigate a jamming-aided wireless edge Metaverse access system to determine the bandwidth required for normal-quality access. The contributions of this paper can be summarized as follows.
	\begin{itemize}
		\item We are the first to propose a bandwidth allocation scheme and a targeted advertising strategy for MSPs under budget constraints. An optimal targeted advertising strategy and corresponding high-quality access service profit are derived with the help of the Vidale-Wolfe advertising model \cite{yang2021optimal} and the Hamiltonian function.
	\item  We propose a novel metric called Meta-Immersion to measure the Metaverse user's feelings. The Meta-Immersion indicates the quality of the uplink and downlink between the user and the metaverse, as well as the user's experience in the virtual world.
		\item We investigate a jamming-aided covert wireless edge Metaverse access system. To decide how much bandwidth is needed for normal-quality access, we derive the detection error probability, downlink covert rate (CR), and uplink bit-error rate (BER) under different modulations.		
	\end{itemize}

	\section{System Description and Advertising Model}
\subsection{System Description}
\begin{figure}[t]
	\centering
	\includegraphics[scale=0.24]{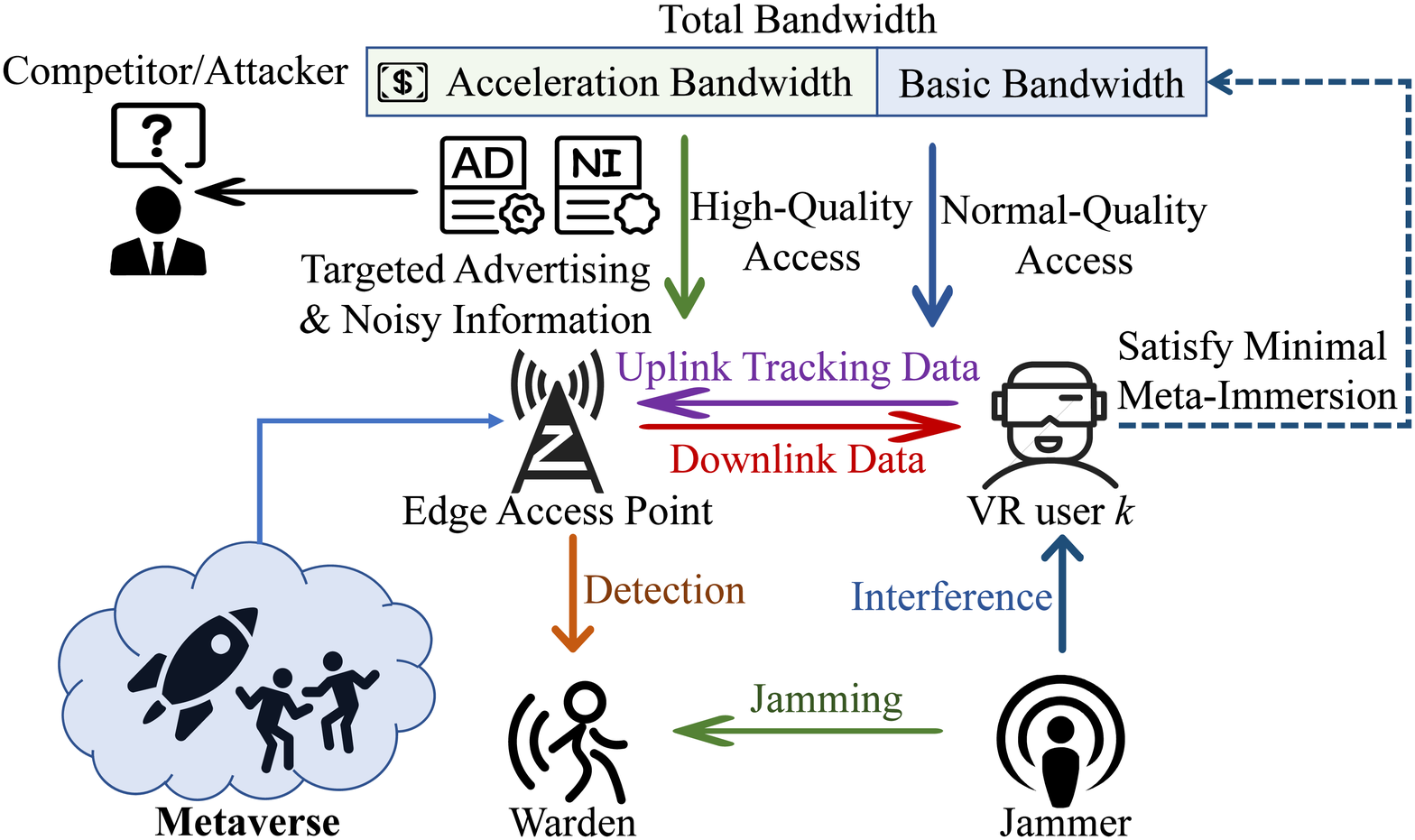}
	\caption{System model of Metaverse covert access and target advertising.}
	\label{model}
\end{figure}
	As shown in Fig. \ref{model}, we consider an MSP with $K$ users, and the VR users accesses to the Metaverse using the head-mounted display (HMD), through EAPs. To prevent the data transmission from being detected by a malicious warden, a friendly jammer is introduced to assist the communication by actively generating jamming signals. For an MSP, the budget is used not only for the construction of virtual worlds, but also for ensuring reliable access for its users. Let $N$ denote the total budget that is planned to be used for Metaverse access over a period of time, i.e., $\left[ {0,T_1} \right]$. The MSP buys $B_T$ bandwidth from the wireless communication providers in unit price $p_l$.

We propose a novel performance metric called Meta-Immersion (more details in Section \ref{MI}) to represent user virtual experiences in Metaverse. Hence, to ensure that each user can achieve the Meta-Immersion requirement, the MSP first provides $B_{k}$ $\left( k=1,\ldots,K\right) $ (called basic bandwidth) to users with a low price for normal-quality access. The $B_{k}$ can be decided by the value of the given minimum Meta-Immersion, \eqref{mi}, and \eqref{rkd}. In addition, some users might want to buy more bandwidth, such as those who prefer to play the Metaverse game at lower latency or those who are more sensitive to cybersickness \cite{venkatakrishnan2020structural}. The MSP can use {\small $\left( {B_T} - \sum\limits_{k = 1}^{N} {{B_k}} \right) $} (called acceleration bandwidth) to provide high-quality access service, but with a higher unit bandwidth price.

\subsection{Covert Advertising Model}
The MSP (MSP A) needs to decide its advertising strategies to attract users to subscribe for the high-quality access service. Let $a\left( t \right)$ denote an actual information advertising effort at time $t$, $e\left( t \right)$ denote the noisy information effort \cite{anand2009targeted}. Note here that the advertising effort can be the square root of budget that the MSP uses to deliver advertising messages to the targeted customers, and the noisy information is cost-free \cite{sethi1983deterministic}.

If a competitor MSP (MSP B) tries to detect the presence of MSP A's advertisement in the market, what Company B faces is a binary decision between the {\it null hypothesis} that MSP A is mute and {\it the non-null hypothesis} that MSP A is advertising. There are two situations where an error in judgment can occur:
{\small \begin{equation}
		{j_{error}}\! = \!\left\{ {\begin{array}{*{20}{l}}
			\!\!\!{\Pr \left( {{\delta _1}a\left( t \right) + {\delta _2}e\left( t \right) < j\left( t \right)} \right),\: {Miss\:Detection},} \\ 
			\!\!\!	{\Pr \left( {{\delta _2}e\left( t \right) > j\left( t \right)} \right),\: {False\:Judgement},}  
				
		\end{array}} \right.
\end{equation}}\noindent
where $j(t)$ is the judgment thresholds, ${\delta _1}$ and ${\delta _2}$ are the message numbers suspected by MSP B to be advertisements under unit advertising and noisy efforts, respectively. Note that MSP A can adjust $e\left( t \right)$ to make ${j_{error}}$ close to one.

Let $B\left( t\right) $ denote the sales of acceleration bandwidth. With the help of a variation of Vidale-Wolfe advertising response model \cite{sethi1983deterministic}, we can express the relationship between the reduction of $B\left( t\right) $ and the targeted advertising effect $a\left( t\right) $ as
{\begin{equation}\label{bdong}
	\frac{{\partial B\left( t \right)}}{{\partial t}} = {c_1}a\left( t \right)\sqrt {1 - \frac{{B\left( t \right)}}{{M}}}  - {\eta _2}B\left( t \right),
\end{equation}}\noindent
where ${c_1}$ is the response constant, $\eta_2$ denotes the decay constant, and $M$ is the saturation level which can be considered as a constant in analysis \cite{gaimon2002optimal}. 

Because the total bandwidth that can be used for high-quality access is limited, in time period ${T}$, we can find a $T_2$ that satisfies
{\small \begin{equation}\label{bandbound}
		\int_0^{{T_2}} {B\left( t \right)dt}  + \sum\limits_{k = 1}^K {{B_k}}  \le {B_T}.
\end{equation}}\noindent
By letting {\small $x\left( t\right)={B\left( t\right) }/{M}$}, {\small $\eta_1={c_1}/{M}$}, we can re-write \eqref{bdong} as
{\small \begin{equation}\label{XD}
		\frac{{\partial x\left( t \right)}}{{\partial t}} = {\eta _1}a\left( t \right)\sqrt {1 - x\left( t \right)}  - {\eta _2}x\left( t \right),
\end{equation}}\noindent
where $x(0)=x_0$. Let $\pi $ denote the maximum acceleration bandwidth sales revenue corresponding to $x\left( t\right) =1$. Recall that the cost of advertising is a quadratic function of the effort \cite{sethi1983deterministic}, the total profit of the MSP can be expressed as
\begin{equation}
	\Pi\triangleq J - {p_l}{{B_T}} \triangleq \int_0^T {\left( {\pi x\left( t \right) - \frac{h_a}{2}a^2\left( t \right)} \right)dt}  - {p_l}{{B_T}},
\end{equation}
where $J$ is the high-quality access service profit, $ T = \min \left\{ {{T_1},{T_2}} \right\} $, and $h_a$ denotes the increasing marginal cost.

\section{Problem Formulation and Analysis}
\subsection{Problem Formulation}
After meeting the basic bandwidth requirements of the $K$ users, the MSP wishes to sell acceleration bandwidth to make more profits. Because the total budget is limited, the budget used in targeted advertising is bounded by
{\small \begin{equation}\label{afanwei}
		0 \le \int_0^T {\frac{{{h_a}}}{2}{a^2}\left( t \right)dt}  \le N- {p_l}{B_T}.
\end{equation}}\noindent
Following the method in \cite{yang2015optimal}, the targeted advertising budget constraint can be replaced by a state variable $G(t)$, defined as
{\small \begin{equation}\label{GY}
		G\left( t \right) = G\left( T \right) + \int_t^T \frac{{{h_a}}}{2}{a^2\left( s \right)ds},
\end{equation}}\noindent
where $ G\left( T \right) \ge 0 $, $G\left( 0 \right) =  N- {p_l}{B_T}$, and 
{\small \begin{equation}\label{GD}
		\frac{{\partial G\left( t \right)}}{{\partial t}} =  - \frac{{{h_a}}}{2}{a^2}\left( t \right).
\end{equation}}\noindent
Because ${p_l}{{B_T}}$ is given, the equivalent optimal control formulation is then given as follows.
{\small \begin{equation}
		\begin{array}{*{20}{c}}
			{\mathop {\max }\limits_{a\left( t \right)} }&J \\ 
			{s.t.}&\eqref{bandbound}, \eqref{XD}, \eqref{afanwei}, \eqref{GY}, \eqref{GD}
		\end{array}
\end{equation}}\noindent
\subsection{Optimal Covert Advertising Solution}
To obtain the optimal targeted advertising strategy, we solve the profit maximization problem with the help of Pontryagin's maximum principle \cite{yang2015optimal}. The current-value Hamiltonian $H$ can be expressed as
{\small \begin{align}
		H =& \pi x\left( t \right) + {\lambda _1}\left(t\right)\left( {{\eta _1}a\left( t \right)\sqrt {1 - x\left( t \right)}  - {\eta _2}x\left( t \right)} \right)
		\notag\\&
		- \frac{{{h_a}}}{2}{a^2}\left( t \right) - {\lambda _2}\left(t\right)\frac{{{h_a}}}{2}{a^2}\left( t \right),
\end{align}}\noindent
where $\lambda_1\left(t\right)$ and $\lambda_2\left(t\right)$ denote the shadow prices, and the dynamic satisfies the following adjoint equations \cite{yang2015optimal}
{\small \begin{subequations}\label{subeq}
		\begin{align}
			\frac{{\partial {\lambda _1}\left( t \right)}}{{\partial t}} &=  - \frac{{\partial H}}{{\partial x}} =  - \pi  + \frac{{{\lambda _1}{\eta _1}e\left( t \right)}}{{2\sqrt {1 - x\left( t \right)} }} + {\lambda _1}{\eta _2} ,\\
			\frac{{\partial {\lambda _2}\left( t \right)}}{{\partial t}} &=  - \frac{{\partial H}}{{\partial G}} = 0 \label{lan2},
		\end{align}
\end{subequations}}\noindent
where ${\partial {\lambda _2}\left( T \right)}\ge0$. Considering the advertising budget constraints, i.e., \eqref{afanwei}, we can solve that ${\lambda _2}=C_2$, where $C_2>0$. The relationship between $C_2$ and $\left( N- {p_l}{B_T}\right) $ will be discussed later.
\begin{them}\label{them1}
	The optimal targeted advertising strategy and corresponding high-quality access service profit can be derived as
	{\small \begin{equation}\label{afeed}
			{a^*}\left( t \right) =  \frac{{{\bar\lambda _1}{\eta _1}\sqrt {1 - x\left( t \right)} }}{{{C_2}{h_a} + {h_a}}},
	\end{equation}}\noindent
{\small \begin{align}\label{optimalJ}
		{J^*} \!= \!\bar x\left(\! {\pi \! + \!\frac{{\Lambda {{\bar\lambda} _1}}}{{2\left(\! {1\! +\! {C_2}} \!\right)}}} \!\right)\!\left(\! {T \!+\! \frac{{{x_0} \!-\! \bar x}}{\Lambda }\!\left(\! {1\! -\! {e^{ - \frac{{\Lambda T}}{{\bar x}}}}} \!\right)}\!\right) \!-\! \frac{{\Lambda {{\bar\lambda} _1}T}}{{2\left(\! {1 \!+\! {C_2}} \!\right)}},
\end{align}}\noindent
where {\small $ \Lambda  \buildrel \Delta \over = \frac{{{{\bar \lambda} _1}{\eta _1}^2}}{{\left( {1 + {C_2}} \right){h_a}}} $}, and 
{\small \begin{subequations}\label{longrun}
		\begin{align}
			{{\bar \lambda }_1} &= \frac{{\sqrt {{\eta _2}^2 + \frac{{2\pi {\eta _1}^2}}{{{C_2}{h_a} + {h_a}}}}  - {\eta _2}}}{{{\eta _1}^2{{\left( {{C_2}{h_a} + {h_a}} \right)}^{ - 1}}}},\\
			\bar x &= \frac{{{{\bar \lambda }_1}{\eta _1}^2}}{{{{\bar \lambda }_1}{\eta _1}^2 + {\eta _2}\left( {{C_2}{h_a} + {h_a}} \right)}}.
		\end{align}
\end{subequations}}\noindent
\end{them}
\begin{IEEEproof}
By differentiating the Hamiltonian $H$ with respect to $a(t)$, we obtain
{\small \begin{equation}
		\frac{{\partial H}}{{\partial a\left( t \right)}} = {\lambda _1}{\eta _1}\sqrt {1 - x\left( t \right)}  - \left( {{C_2}{h_a} + {h_a}} \right)a\left( t \right).
\end{equation}}\noindent
Thus, the optimal targeted advertising management strategy in feedback form can be obtained as \eqref{afeed}. Substituting \eqref{afeed} into \eqref{subeq} and \eqref{XD}, and letting $ \frac{{\partial {\lambda _1}\left( t \right)}}{{\partial t}} =\frac{{\partial x\left( t \right)}}{{\partial t}}=0$, we can obtain the optimal long-run stationary equilibrium \cite{sethi1983deterministic} as \eqref{longrun}, and $ {{a^*}\left( t \right)} $ can be derived by substituting \eqref{longrun} into \eqref{afeed}. To obtain the optimal paths for $x$ and ${\lambda }_1$ for any given value of $x_0$, we set $ {\lambda _1}\left( 0 \right) = {{\bar \lambda }_1} $ in \eqref{subeq} and \eqref{XD} \cite{sethi1983deterministic}. Thus, we can obtain
{\small \begin{subequations}\label{solution}
		\begin{align}
			x\left( t \right) &= \left( {{x_0} - \bar x} \right){e^{ - \left( {\frac{{{{\bar\lambda} _1}{\eta _1}^2}}{{{h_a} + {C_2}{h_a}}}\frac{1}{{\bar x}}} \right)t}} + \bar x,  \\
			{\lambda _1}\left( t \right) &= {{\bar \lambda }_1}.
		\end{align}
\end{subequations}}\noindent
Using the definition of $J$ and \eqref{afeed}, we have
{\small \begin{equation}\label{jstar}
		{J^*} \!=\!\! \int_0^T \!\!{\left(\! {\left(\! {\pi \! +\! \frac{1}{2}\frac{{{{\bar \lambda }_1}^2{\eta _1}^2}}{{{C_2}^2{h_a} \!+ \!{h_a}}}} \!\right)\!x\left( t \right) \!-\! \frac{1}{2}\frac{{{{\bar \lambda }_1}^2{\eta _1}^2}}{{{C_2}^2{h_a}\! + \!{h_a}}}} \!\right)\!dt}.
\end{equation}}\noindent
Substituting \eqref{solution} into \eqref{jstar}, after some algebraic manipulations, we can solve ${J^*}$ as \eqref{optimalJ} to complete the proof.
\end{IEEEproof}
\begin{rem}
	Substituting \eqref{solution} into \eqref{afanwei}, we can obtain that $C_2$ must satisfy
	{\small \begin{equation}
			{C_2} \ge\! \frac{{T{{\bar \lambda }_1}\Lambda  + {{\bar \lambda }_1}\bar x\left(\! {\Lambda T \!+\! \left(\! {{x_0} \!- \!\bar x} \!\right)\!\!\left(\! {1 \!-\! {e^{ - \frac{{\Lambda T}}{{\bar x}}}}}\! \right)} \!\right)}}{{2\left( {M - {p_l}{B_T}} \right)}}\! - \!1.
	\end{equation}}\noindent
In practical situation where the MSP has a limited advertising budget, $N- {p_l}{B_T}$, the detailed steps for finding the ${C_2}$ and optimal advertising budget allocation strategy are similar to \cite[Algorithm 1]{yang2021optimal}.
\end{rem}
\begin{rem}
By substituting \eqref{solution} into \eqref{bandbound}, we have
{\small \begin{equation}
		f\left( {{T_2}} \right)\!\triangleq\!{T_2} -\! \frac{{{x_0} - \bar x}}{\Lambda }{e^{ - \frac{{\Lambda {T_2}}}{{\bar x}}}}\! \le \!\frac{{{B_T} \!- \!\sum\limits_{k = 1}^N {{B_k}} }}{{\bar xM}} - \frac{{{x_0} - \bar x}}{\Lambda }.
\end{equation}}\noindent
Note that $f\left( {{T_2}} \right)$ is a monotonically increasing function, we observe that increasing $ {\sum\limits_{k = 1}^N {{B_k}} } $ will decrease $T_2$, which potentially results in a decrease in ${J^*}$. Therefore, it is reasonable for the MSP to allocate as little basic bandwidth as possible while satisfying the user's minimal Meta-Immersion.
\end{rem}
\section{Meta-Immersion of Users}\label{MI}
\begin{figure}[t]
	\centering
	\includegraphics[scale=0.27]{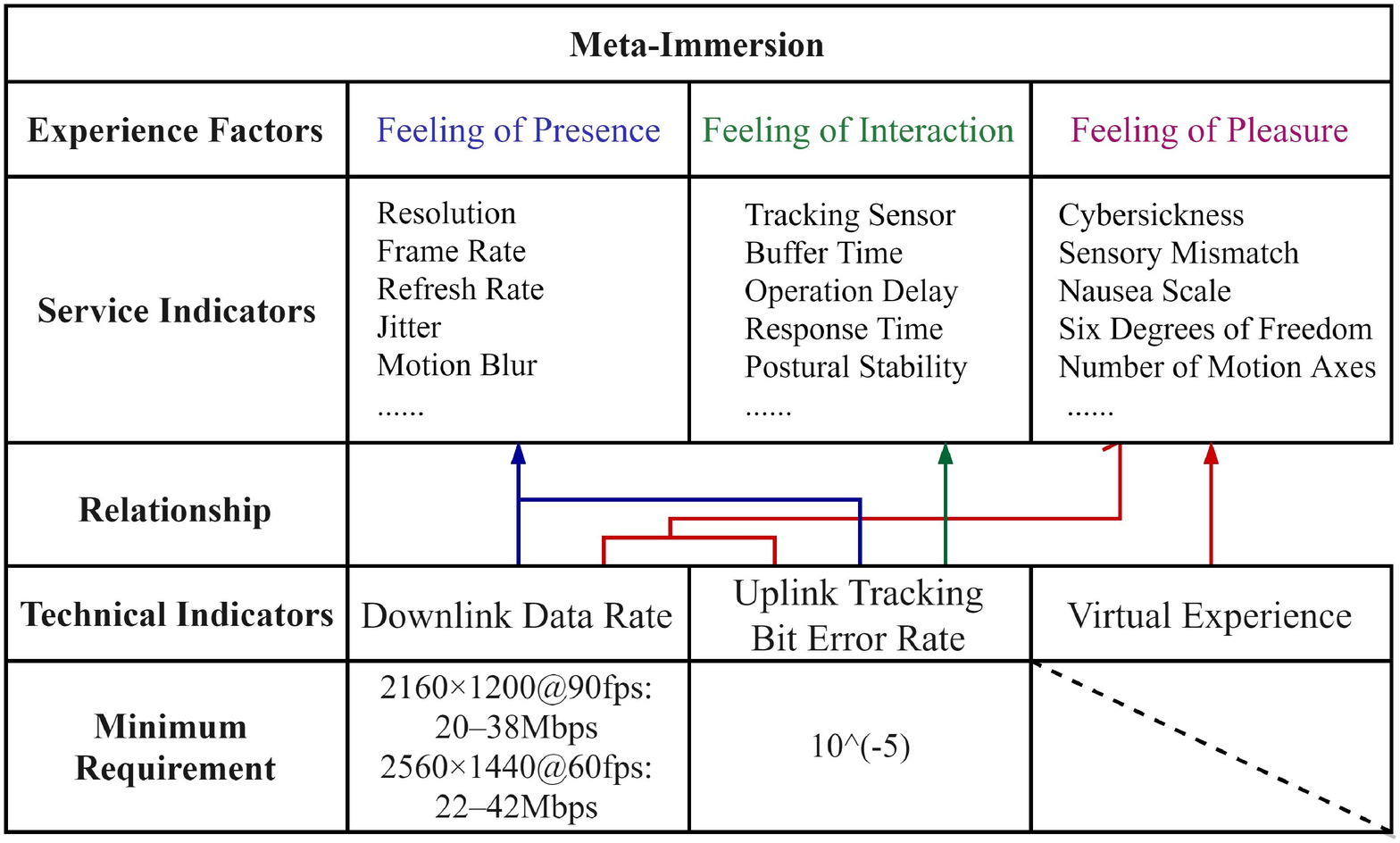}
	\caption{A novel performance metric in Metaverse: Meta-Immersion, and coresponding experience factors, service indicators, technical indicators and minimum requirements.}
	\label{NemI}
\end{figure}

To determine the basic bandwidth that an MSP should allocate to its users, we first propose a novel metric to represent the user feelings in the Metaverse. We divide the users' experience and service indicators into three groups, shown as Fig. \ref{NemI}. 
The technical indicators are discussed as follows:
\begin{itemize}
	\item {\it Downlink Data Rate $\left(R^d_{k}\right) $}: Downlink data rate should be high enough to provide a visually lossless experience. According to \cite{wang2004image}, the minimum rate is given in Fig. \ref{NemI}.
	\item {\it Uplink Tracking Bit Error Rate $\left( E^u_k\right) $}: According to \cite{boardieee}, the minimum packet error rate is $10^{-2}$. By considering that the packet length is $10^3$ bit \cite{ieee802ieee}, a recommended BER could be $10^{-5}$.
	\item {\it Virtual Experience $\left( S_k\right) $}: This indicator depends on the subjective behavior of the user, such as activity in the Metaverse, the length of online time, the user's physical fitness, etc. Quantifying of $S_k$ can be done with the help of Structural Equation Modeling (SEM).
\end{itemize}
Therefore, the Meta-Immersion for user $k$, $M\!I_k$, can be defined as 
\begin{equation}\label{mi}
	{M\!I}_k=R^d_{k}(1-E^u_k)S_k.
\end{equation}
 Because the main contribution of this paper is to investigate the Metaverse covert access, we consider $S_k$ is fixed. A EAP transmits the data signals to user $k$, and the downlink bandwidth is $B_k$. In the following, we derive the warden's DEP ${\xi _w}$, $R^d_{k}$, and $ E^u_k$.
\subsection{Channel Model}
We consider that jamming links follow the $ \alpha  - \mu  $ distribution, which includes several important distributions such as the One-Sided Gaussian, Rayleigh (when the jamming channels are frequency-flat), Nakagami-$m$ (when the jamming channels are frequency-selective) and Weibull \cite{yacoub2007alpha}. Let {\small $ \Upsilon  \sim \alpha \mu \left( {\alpha ,\mu ,\bar \gamma } \right) $}, where $ \Upsilon$ is a squared $ \alpha  - \mu  $ RV. The PDF of $ \Upsilon$ is given by
{\small \begin{equation}\label{auPDF}
		{f_\Upsilon }\left( \gamma  \right) = \frac{{\alpha {\gamma ^{\frac{{\alpha \mu }}{2} - 1}}}}{{2{\beta ^{\frac{{\alpha \mu }}{2} }}\Gamma \left( \mu  \right)}}\exp \left( { - {{\left( {\frac{\gamma }{\beta }} \right)}^{\frac{\alpha }{2}}}} \right),
\end{equation}}\noindent
where {\small $\Gamma\left(\cdot \right) $} is the gamma function \cite[eq. (8.310.1)]{gradshteyn2007}, {\small $ \beta  = \frac{{\bar \Upsilon \Gamma \left( \mu  \right)}}{{\Gamma \left( {\mu  + \frac{2}{\alpha }} \right)}} $} and {\small ${\bar \gamma}=E\left(\gamma \right) $}.
The CDF is given as
{\small \begin{equation}\label{cdfau}
		{F_\Upsilon }\left( \gamma  \right) = \frac{{\gamma  \left( {\mu ,{\gamma ^{\frac{\alpha }{2}}}{\beta ^{ - \frac{\alpha }{2}}}} \right)}}{{\Gamma \left( \mu  \right)}},
\end{equation}}\noindent
where $\gamma\left(\cdot \right) $ is the incomplete gamma function \cite[eq. (8.35)]{gradshteyn2007}.

We consider that data links follow the Fisher-Snedecor $\mathcal{F}$ fading, which can give an accurate modeling of the simultaneous occurrence of multi-path fading and shadowing, and covers several fading models. Let {\small $ Z\sim\mathcal{F}\left( {{m},{m_s},\bar \kappa} \right) $}, the PDF and CDF of the squared $\mathcal{F}$ RV $Z$ can be written as \cite{yoo2019comprehensive}
{\small \begin{equation}
		{f_Z}\left( z \right) = \frac{{{m}^{{m}}{{\left( {{m_s} - 1} \right)}^{{m_s}}}{{\bar \kappa}^{{m_s}}}{z^{{m} - 1}}}}{{B\left( {{m},{m_s}} \right){{\left( {{m}z + \left( {{m_s} - 1} \right)\bar \kappa} \right)}^{{m} + {m_s}}}}},
\end{equation}}\noindent
{\small \begin{align}\label{cdff}
		{F_Z}\left( z \right) = \frac{{{z^m}_2{F_1}\left( {m,m + {m_s},m + 1; - \frac{{mz}}{{\left( {{m_s} - 1} \right)\bar \kappa }}} \right)}}{{{m^{1 - m}}B\left( {m,{m_s}} \right)\left( {{m_s} - 1} \right){{\bar \kappa }^m}}},
\end{align}}\noindent
where ${}_2F_1(\cdot,\cdot,\cdot;\cdot)$ is the Gauss hypergeometric function \cite[eq. (9.111)]{gradshteyn2007}, and $B(\cdot,\cdot)$ is the Beta function \cite[eq. (8.384.1)]{gradshteyn2007}.
\subsection{Detection Error Probability Analysis}
The DEP of the warden, ${\xi _w}$, can be defined as the sum of the false alarm probability ({\small $ {\Pr \left( {{p_j}{{\left| {{h_{jw}}} \right|}^2} \!+\! \sigma _{aw}^2 > \varepsilon } \right)} $}) and the miss detection probability ({\small$  {\Pr \left( {{p_a}{{\left| {{h_{aw}}} \right|}^2} + {p_j}{{\left| {{h_{jw}}} \right|}^2} + \sigma _{aw}^2 < \varepsilon } \right)} $}) \cite{zheng2019multi}, where $\sigma _{aw}^2$ is the noise power, $p_a$ is the transmit power of EAP, $p_j$ is the jamming power, $\varepsilon$ is the detection threshold, {\small $ {\left| {{h_{jw}}} \right|^2} \sim \alpha \mu \left( {{\alpha _{jw}},{\mu _{jw}},{{\bar \gamma }_{jw}}} \right) $}, and {\small $ {\left| {{h_{aw}}} \right|^2} \sim {\cal F}\left( {{m_{aw}},{m_{{s_{aw}}}},{{\bar \kappa }_{aw}}} \right) $}. 
\begin{them}\label{AppendixAA}
	The close-form of ${\xi _w}$ can be derived as \eqref{derror}, shown at the top of the next page.
\end{them}
\newcounter{mytempeqncnt3H}
\begin{figure*}[t]
	\normalsize
	\setcounter{mytempeqncnt3H}{\value{equation}}
	\setcounter{equation}{24}
	{\footnotesize \begin{align}\label{derror}
			{\xi _w} =& 1-\frac{1}{{\Gamma \left( {{\mu _{jw}}} \right)}}\gamma \left( {{\mu _{jw}},{{\left( {\frac{{\varepsilon  - \sigma _{aw}^2}}{{{p_j}}}} \right)}^{\frac{{{\alpha _{jw}}}}{2}}}{\beta _{jw}}^{ - \frac{{{\alpha _{jw}}}}{2}}} \right) + \frac{{{m_{aw}}^{{m_{aw}}}{{\left( {{m_{{s_{aw}}}} - 1} \right)}^{{m_{{s_{aw}}}}}}{{\bar \kappa }_{aw}}^{{m_{{s_{aw}}}}}{{\left( {\varepsilon  - \sigma _{aw}^2} \right)}^{{m_{aw}}}}}}{{\Gamma \left( {{\mu _{jw}}} \right)\Gamma \left( {{m_{aw}}} \right)\Gamma \left( {{m_{{s_{aw}}}}} \right){p_a}^{{m_{aw}}}{{\left( {\left( {{m_{{s_{aw}}}} - 1} \right){{\bar \kappa }_{aw}}} \right)}^{{m_{aw}} + {m_{{s_{aw}}}}}}}}
			\notag\\&\times
		H_{0,1:2:2;2,1}^{0,0:1,2;1,2}\left( {\left. {\begin{array}{*{20}{c}}
					{{{\left( {\frac{{{p_j}{\beta _{jw}}}}{{\varepsilon  - \sigma _{aw}^2}}} \right)}^{ - \frac{{{\alpha _{jw}}}}{2}}}}\\
					{\frac{{{m_{aw}}\left( {\varepsilon  - \sigma _{aw}^2} \right)}}{{{p_a}\left( {{m_{{s_{aw}}}} - 1} \right){{\bar \kappa }_{aw}}}}}
			\end{array}} \right|\begin{array}{*{20}{c}}
				{ - :\left( {1,1} \right)\left( {0,\frac{{{\alpha _{jw}}}}{2}} \right);\left( {1 - {m_{aw}} - {m_{{s_{aw}}}},1} \right)\left( {1 - {m_{aw}},1} \right)}\\
				{\left( { - {m_{aw}};\frac{{{s_1}{\alpha _{jw}}}}{2},1} \right):\left( {{\mu _{jw}},1} \right)\left( {0,1} \right);\left( {0,1} \right)}
		\end{array}} \right)
	\end{align}}
	\setcounter{equation}{\value{mytempeqncnt3H}}
	\hrulefill
\end{figure*}
\setcounter{equation}{25}
\begin{IEEEproof}
	Let {\small $ {Y_1} = {p_j}{\left| {{h_{jw}}} \right|^2} $}. We have {\small ${F_{{Y_1}}}\left( y \right) =\! {F_{{{\left| {{h_{jw}}} \right|}^2}}}\!\left(\! {\frac{{y }}{{{p_j}}}} \!\right)$}. Thus, the false alarm probability can be derived as {\small $\Pr \left( {{p_j}{{\left| {{h_{jw}}} \right|}^2}  > \varepsilon\!- \sigma _{aw}^2 } \right) = 1 - {F_{{Y_1}}}\left( \varepsilon -\sigma _{aw}^2  \right)$}. Let ${Y_2} = {p_a}{\left| {{h_{aw}}} \right|^2} + {Y_1}$, the CDF of $Y_2$ can be expressed as {\small ${F_{{Y_2}}}\left( y \right)= \int_0^\infty  {{F_{{Y_1}}}\left( {y - t} \right)\frac{1}{{{p_a}}}{f_{{{\left| {{h_{aw}}} \right|}^2}}}\left( {\frac{t}{{{p_a}}}} \right){\rm{d}}t}$}. Substituting ${F_{{Y_1}}}$ and \eqref{auPDF} into ${F_{{Y_2}}}$, we have
	{\small \begin{equation}
			{F_{{Y_2}}}\left( y \right) = \frac{{{m_{aw}}^{{m_{aw}}}{{\left( {{m_{{s_{aw}}}} - 1} \right)}^{{m_{{s_{aw}}}}}}{{\bar \kappa }_{aw}}^{{m_{{s_{aw}}}}}}}{{{p_a}^{{m_{aw}}}\Gamma \left( {{\mu _{jw}}} \right)B\left( {{m_{aw}},{m_{{s_{aw}}}}} \right)}}{I_{{A_1}}},
	\end{equation}}\noindent
	where
	{\small \begin{equation}
			{I_{{A_1}}} = \int_0^\infty  {\frac{{\gamma \left( {{\mu _{jw}},{{\left( {\frac{{y - t }}{{{p_j}}}} \right)}^{\frac{{{\alpha _{jw}}}}{2}}}{\beta _{jw}}^{ - \frac{{{\alpha _{jw}}}}{2}}} \right){t^{{m_{aw}} - 1}}}}{{{{\left( {\frac{{{m_{aw}}t}}{{{p_a}}} + \left( {{m_{{s_{aw}}}} - 1} \right){{\bar \kappa }_{aw}}} \right)}^{{m_{aw}} + {m_{{s_{aw}}}}}}}}{\rm{d}}t} .
	\end{equation}}\noindent
	With the help of \cite[eq. (06.06.07.0005.01)]{web}, \cite[eq. (2.2.6.15)]{Prudnikov1986Integrals}, \cite[eq. (8.384.1)]{gradshteyn2007}, and \cite[eq. (9.113)]{gradshteyn2007}, $I_{A1}$ can be solved. Substituting $I_{A1}$ into ${F_{{Y_2}}}$, we can obtain the miss detection probability as ${\small {F_{{Y_2}}}\left( \varepsilon - \sigma _{aw}^2 \right)}$ to complete the proof.
\end{IEEEproof}
\subsubsection{Covert Rate Analysis}
When the DEP is close to $1$, the CR is defined as
{\small \begin{equation}\label{covertrate}
		R_k^d = {B_k}{\log _2}\left( {1 + \frac{{{p_a}{{\left| {{h_{ak}}} \right|}^2}}}{{{B_k}\sigma _{ak}^2 + {p_j}{{\left| {{h_{jk}}} \right|}^2}}}} \right),
\end{equation}}\noindent
where {\small $ {\left| {{h_{ak}}} \right|^2}\! \sim \!{\cal F}\left( {{m_{ak}},{m_{{s_{ak}}}},{{\bar \kappa }_{ak}}} \right) $}, {\small $ {\left| {{h_{jk}}} \right|^2}\! \sim \!\alpha \mu \left( {{\alpha _{jk}},{\mu _{jk}},{{\bar \gamma }_{jk}}} \right) $}, {\small $\sigma _{ak}^2=-174$ ${\rm dBm/Hz}$}, and $B_k$ is the downlink bandwidth.
\begin{them}\label{AppendixBB}
	The close-form of $R_k^d$ can be derived as \eqref{rkd}, which is shown at the top of the next page.
\end{them}
\newcounter{mytempeqncnt1}
\begin{figure*}[t]
	\normalsize
	\setcounter{mytempeqncnt1}{\value{equation}}
	\setcounter{equation}{28}
	{\footnotesize \begin{align}\label{rkd}
			R_k^d =\! \frac{{{\alpha _{jk}}}}{{\ln4\Gamma \!\left( {{\mu _{jk}}} \right)\Gamma \!\left( {{m_{ak}}} \right)\Gamma \!\left( {{m_{{s_{ak}}}}} \right)}}
		H_{2,0:0,2;0,2;2,3}^{0,2:2,0;1,0;3,1}\!\!\left(\!\!\!\! {\left. {\begin{array}{*{20}{c}}
					{\frac{{{B_k}\sigma _{ak}^2}}{{{\beta _{jk}}{p_j}}}}\\
					{ - 1}\\
					{\frac{{{m_{ak}}{B_k}\sigma _{ak}^2}}{{{p_a}\left( {{m_{{s_{ak}}}} - 1} \right){{\bar \kappa }_{ak}}}}}
			\end{array}} \!\!\!\right|\!\!\!\begin{array}{*{20}{c}}
				{\left( {1;1,1,0} \right)\left( {{m_{{s_{ak}}}};0,1,1} \right)\!:\! - ;\! - ;\left( {0,1} \right)\left( {1,1} \right)}\\
				{ - \!:\!\left( {{m_{{s_{ak}}}},1} \right)\left( {{\mu _{jk}},\frac{2}{{{\alpha _{jk}}}}} \right)\!;\!\left( {0,1} \right)\left( {1 - {m_{{s_{ak}}}},1} \right);\left( {0,1} \right)\left( {0,1} \right)\left( {{m_{ak}},1} \right)}
		\end{array}} \!\!\!\!\right)
	\end{align}}
	\setcounter{equation}{\value{mytempeqncnt1}}
	\hrulefill
\end{figure*}
\setcounter{equation}{29}
\begin{IEEEproof}
	Let {\small $ {X_1} = {p_a}{\left| {{h_{ak}}} \right|^2} $}, {\small $ {X_2} = {B_k}\sigma _{ak}^2 + {p_j}{\left| {{h_{jk}}} \right|^2} $} and {\small $X={X_1}/{X_2}$}. We have {\small ${f_{{X_1}}}\left( x \right) = \frac{1}{{{p_a}}}{f_{{{\left| {{h_{ak}}} \right|}^2}}}\left( {\frac{x}{{{p_a}}}} \right)$} and {\small ${f_{{X_2}}}\left( x \right) = \frac{1}{{{p_j}}}{f_{{{\left| {{h_{jk}}} \right|}^2}}}\left( {\frac{{x - {B_k}\sigma _{ak}^2}}{{{p_j}}}} \right)$}. Thus, the PDF of $X$ can be expressed as {\small $ {f_X}(x) = \int_0^\infty  y {f_{{X_1}}}(xy){f_{{X_2}}}(y){\text{dy}} $}. We only need to solve that 
	{\small \begin{equation}
			{I_{{B_1}}} =\int_{{B_k}\sigma _{ak}^2}^\infty  {\frac{{{y^{{m_{ak}}}}{{\left( {y - {B_k}\sigma _{ak}^2} \right)}^{\frac{{{s_1}{\alpha _{jk}}}}{2} + \frac{{{\alpha _{jk}}{\mu _{jk}}}}{2} - 1}}}}{{{{\left( {{m_{ak}}xy + {p_a}\left( {{m_{{s_{ak}}}} - 1} \right){{\bar \kappa }_{ak}}} \right)}^{{m_{ak}} + {m_{{s_{ak}}}}}}}}} {\rm{dy}}.
	\end{equation}}\noindent
	With the help of \cite[eq. (8.384.1)]{gradshteyn2007}, \cite[eq. (9.113)]{gradshteyn2007} and \cite[eq. (3.197.2)]{gradshteyn2007}, $I_{B_1}$ can be solved. By substituting $I_{B_1}$ into $f_X(x)$ and using \cite[eq. (A-1)]{mathai2009h}, $f_X(x)$ can be derived. Substituting $f_X(x)$ into {\small $ R_k^d = {B_k}\int_0^\infty  {{{\log }_2}\left( {1 + \gamma } \right){f_X}\left( \gamma  \right){\text{d}}\gamma }  $}, we only need to solve
	{\small \begin{equation}
			{I_{{B_2}}}\! = \!\!\int_0^\infty \!\!\!\!\!\! {\frac{{\log \left( {1 + x} \right){x^{{m_{ak}} - 1}}}}{{{{\left( {{m_{ak}}x{B_k}\sigma _{ak}^2 \!+\! {p_a}\!\left( {{m_{{s_{ak}}}} \!-\! 1} \right){{\bar \kappa }_{ak}}} \right)}^{{s_2} + {m_{ak}} + {m_{{s_{ak}}}}}}}}dx} .
	\end{equation}}\noindent
	With the help of \cite[eq. (01.04.07.0003.01)]{web} and \cite[eq. (3.194.3)]{gradshteyn2007}, $I_{B_2}$ can be solved. Using ${I_{B_2}}$ and \cite[eq. (A-1)]{mathai2009h}, we can obtain $ R_k^d $ as \eqref{rkd}, which completes the proof.
\end{IEEEproof}
\subsection{Bit Error Rate Analysis}
We consider that jamming signals' interference to the EAP can be neglected, because the frequency division multiplexing is typically used in the downlink and uplink. The signal-to-noise ratio can be expressed as
{\small \begin{equation}\label{gammaka}
		{\gamma _{ka}} = \frac{{{p_{k}}{{\left| {{h_{ka}}} \right|}^2}}}{{\sigma _{ka}^2}},
\end{equation}}\noindent
where {\small $ {\left| {{h_{ka}}} \right|^2}\sim {\mathcal F}\left( {{m_{ka}},{m_{{s_{ka}}}},{{\bar \kappa }_{ka}}} \right) $}, {\small ${{\sigma _{ka}^2}}$} is the noise power in the uplink, and $p_k$ is the transmit power of HMD. For a variety of modulation formats, the average BER, $E_k^u$, is given by 
{\small \begin{equation}
		E_k^u = \int_0^\infty  {\frac{{\Gamma \left( {{\tau _2},{\tau _1}\gamma } \right)}}{{2\Gamma \left( {{\tau _2}} \right)}}{f_{{\gamma _{ka}}}}\left( \gamma  \right)d\gamma },
\end{equation}}\noindent
where $ {{{\Gamma \left( {{\tau _2},{\tau _1}\gamma } \right)}}/{{2\Gamma \left( {{\tau _2}} \right)}}} $ is the conditional bit-error probability, $\Gamma \left( { \cdot , \cdot } \right)$ is the upper incomplete Gamma function \cite[eq. (8.350.2)]{gradshteyn2007}, ${\tau _1} $ and ${\tau _2}$ are modulation-specific parameters for various modulation/detection combinations. For example, $\left\{ {{\tau _1} = 0.5,{\tau _2} = 0.5} \right\}$ is orthogonal coherent binary frequency-shift keying (BFSK), $\left\{ {{\tau _1} = 1,{\tau _2} = 0.5} \right\}$ is antipodal coherent binary phase-shift keying (BPSK), $\left\{ {{\tau _1} = 0.5,{\tau _2} = 1} \right\}$ is orthogonal non-coherent BFSK, and $\left\{ {{\tau _1} = 1,{\tau _2} = 1} \right\}$ is antipodal differentially coherent BPSK (DPSK).
\begin{them}\label{AppendixCC}
	The close-form of $E_k^u$ is derived as \eqref{eku}.
	{\small \begin{align}\label{eku}
			&	E_k^u = \frac{{{m_{sk}}{m_{ka}}^{{m_{ka}} - 1}\sigma {{_{ka}^2}^{{m_{ka}}}}{\tau _1}^{ - {m_{ka}}}}}{{2\Gamma \left( {{\tau _2}} \right){\Gamma ^2}\left( {{m_{ka}}} \right)\left( {{m_{{s_{ka}}}} - 1} \right){{\left( {{p_k}{{\bar \kappa }_{ka}}} \right)}^{{m_{ka}}}}}}\notag
			\\&\!\!\times\!\!
			E\!\!\left(\!\! {{m_{ka}}\!,\!{m_{ka}} \!+ \!{m_{{s_{ka}}}}\!,\!{\tau _2}\! + \!{m_{ka}}\!\!:\!\!{m_{sk}}\! \!+\! 1\!\!:\!\!\frac{{{\tau _1}\!\left(\! {{m_{{s_{ka}}}}\! \!-\! 1} \!\right){{\bar \kappa }_{ka}}}}{{{p_k}^{ - 1}\sigma _{ka}^2{m_{ka}}}}} \!\!\right)\!\!,
	\end{align}}\noindent
where ${\small E\left( { \cdot , \cdot , \cdot : \cdot : \cdot } \right)}$ is the MacRobert's $E$-Function \cite[eq. (9.4)]{gradshteyn2007}.
\end{them}
\begin{IEEEproof}
	Using the definition of Gamma function \cite[eq. (8.350)]{gradshteyn2007}, we can re-write $E_k^u$ as
	{\small \begin{equation}\label{ekudef}
			E_k^u = \frac{{{\tau _1}^{{\tau _2}}}}{{2\Gamma \left( {{\tau _2}} \right)}}\int_0^\infty  {{x^{{\tau _2} - 1}}{e^{ - {\tau _1}x}}{F_{{\gamma _{ka}}}}\left( x \right)d\gamma }.
	\end{equation}}\noindent
	The CDF of $\gamma_{ka}$ can be obtained with the help of \eqref{cdff}. Substituting ${F_{{\gamma _{ka}}}}$ into \eqref{ekudef}, we have
	
	{\small \begin{align}
			E_k^u =& \frac{{{m_{ka}}^{{m_{ka}} - 1}\sigma {{_{ka}^2}^{{m_{ka}}}}}}{{B\left( {{m_{ka}},{m_{{s_{ka}}}}} \right)\left( {{m_{{s_{ka}}}} - 1} \right){{\left( {{p_k}{{\bar \kappa }_{ka}}} \right)}^{{m_{ka}}}}}}
			\notag\\&\times
			\frac{{{\tau _1}^{{\tau _2}}}}{{2\Gamma \left( {{\tau _2}} \right)}}{\left( {\frac{{{p_k}\left( {{m_{{s_{ka}}}} - 1} \right){{\bar \kappa }_{ka}}}}{{\sigma _{ka}^2{m_{ka}}}}} \right)^{{\tau _2} + {m_{ka}}}}{I_C},
	\end{align}}\noindent
	where $I_C$ can be solved with the help of \cite[eq. (7.522.1)]{gradshteyn2007}. Substituting $I_C$ into $E_k^u$ and using \cite[eq. (8.331.1)]{gradshteyn2007}, we can obtain \eqref{eku} to complete the proof.
\end{IEEEproof}
With the help of the derived ${\xi _w}$, $R^d_{k}$, and $ E^u_k$, \eqref{mi} can be expressed as a function of the basic bandwidth. Thus, Given minimum ${M\!I}_k$, the MSP can determine the corresponding $B_k$.
\section{Numerical Results}
\begin{figure*}
	\setcaptionwidth{2.25in}
	\setlength{\abovecaptionskip}{-2pt}
	\setlength{\belowcaptionskip}{-1pt}
	\centering
	\begin{minipage}[t]{0.331\linewidth} 
		\centering
		\includegraphics[width=2.3in]{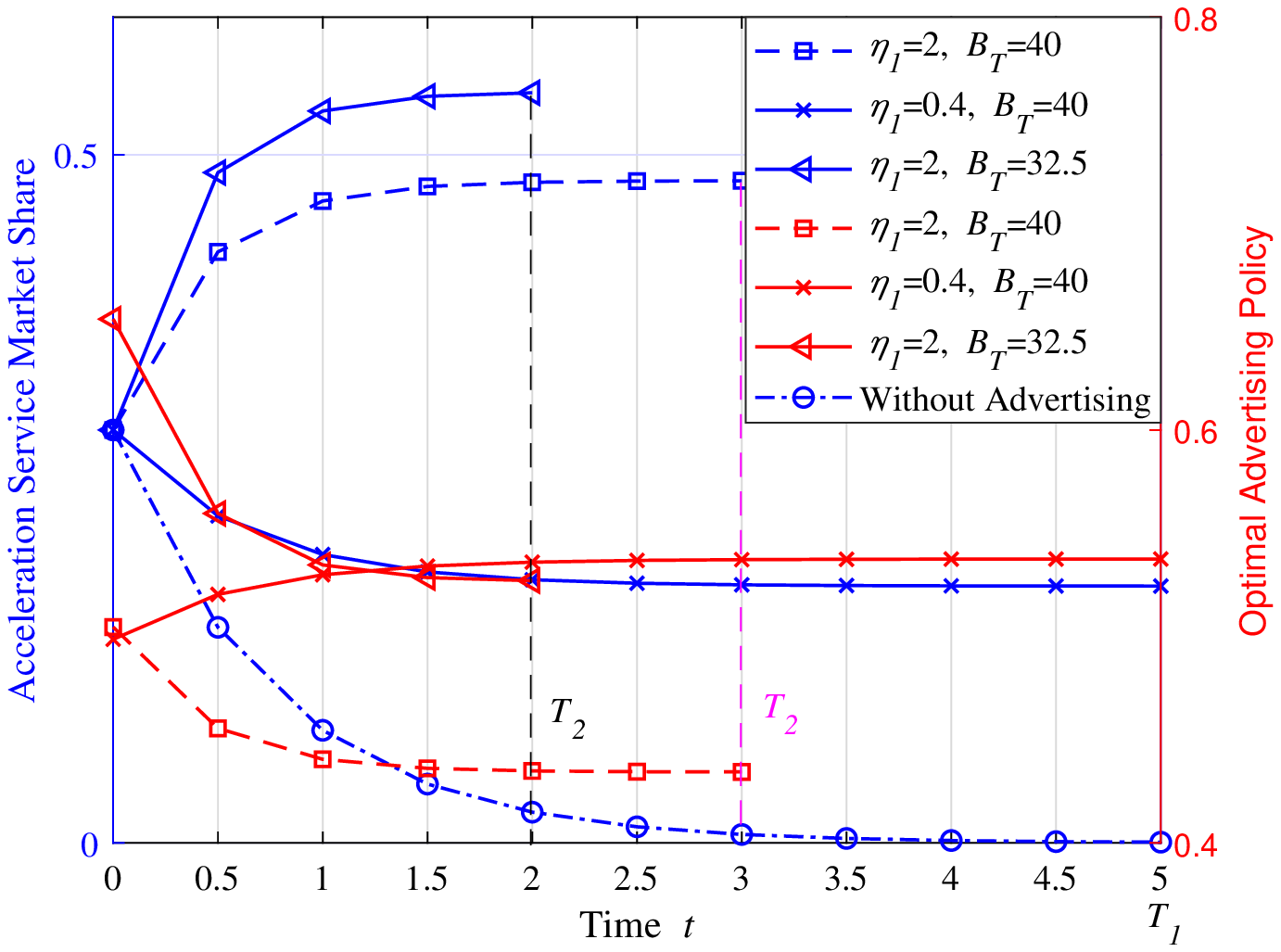}
		\caption{The optimal targeted advertising strategy and high-quality access service market share versus the time, with $ {{h_a}} =3$, $ {\eta _2}=1.3 $, $ \pi=10$, $T_1=5$, $x_0=0.3$, $N=20$, and $p_l=0.4$.}
			\label{fig1}
		\end{minipage}%
		\begin{minipage}[t]{0.331\linewidth}
		\centering
		\includegraphics[width=2.3in]{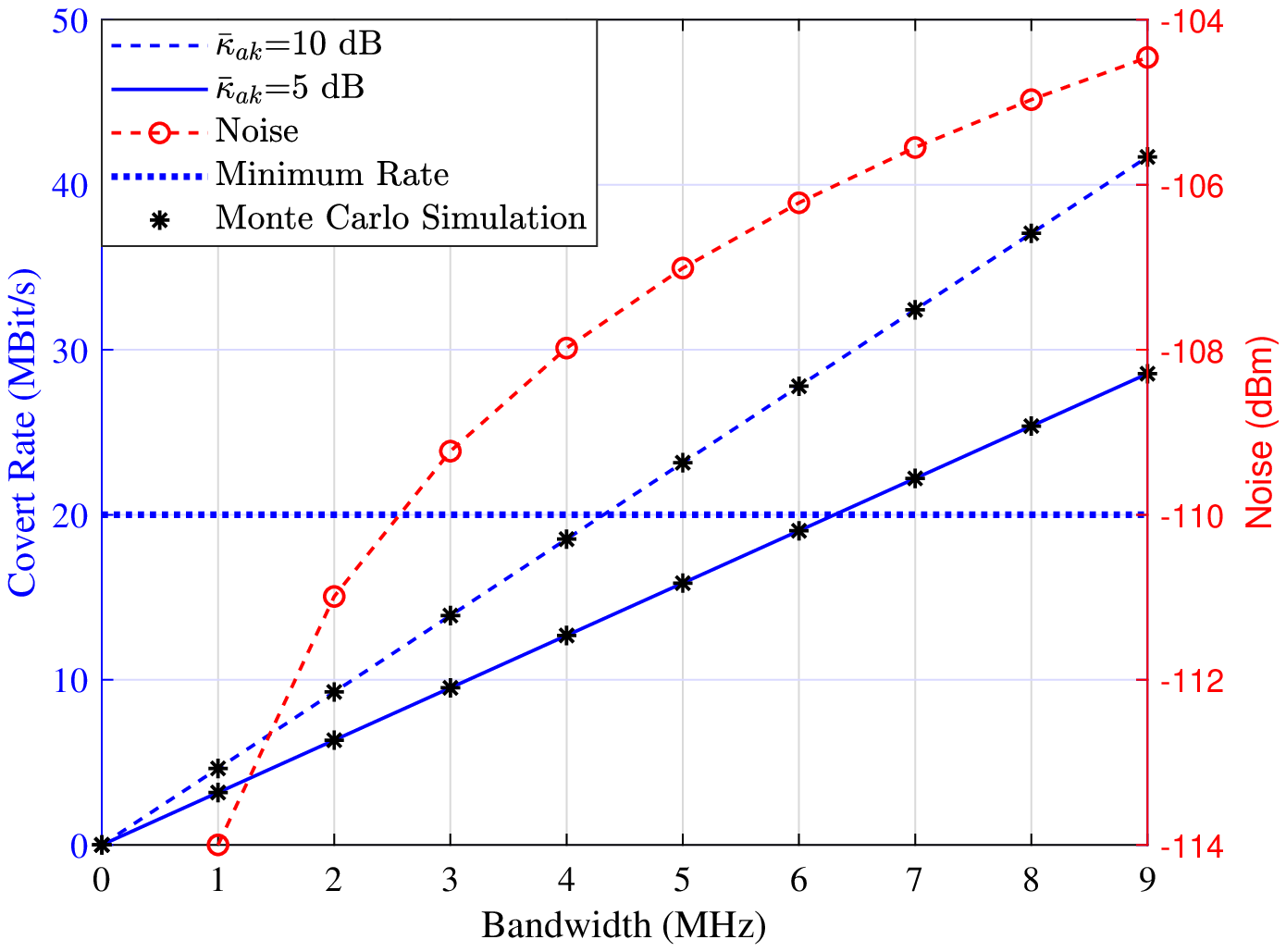}
		\caption{The user's downlink CR and noise versus the allocated bandwidth, with $m_{s_{ak}}=2$, $m_{ak}=3$, ${{\mu _{jk}}}=2$, $ {{\alpha _{jk}}}=2$, $p_a=10$ ${\rm dBW}$, $ {{{\bar \gamma }_{jk}}}=5$ ${\rm {dBW}}$, and $ \delta=3\%$.}
		\label{fig2}
	\end{minipage}
	\begin{minipage}[t]{0.331\linewidth}
		\centering
		\includegraphics[width=2.3in]{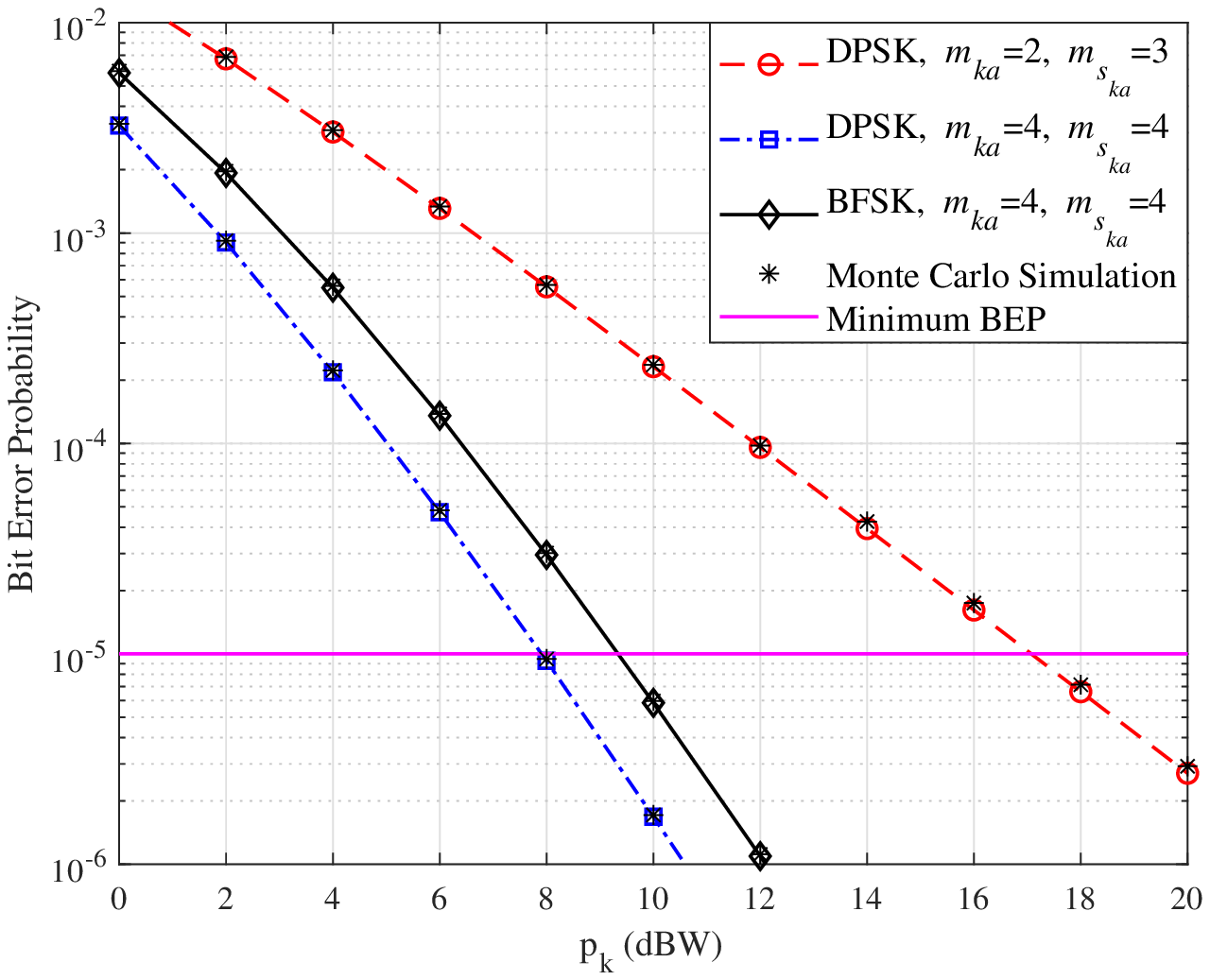}
		\caption{The user's uplink BER versus the transmit power, with $ {{{\bar \kappa }_{ka}}}=10$ ${\rm dBW}$, and $ {\sigma _{ka}^2}=1$ ${\rm dBW}$.}
		\label{fig3}
	\end{minipage}
	\end{figure*}

Simulation results are presented to verify the proposed analysis in this section. We set $ {{h_a}} =3$, $ {\eta _2}=1.3 $, $ \pi=10$, $T_1=5$, $a_U=0.9$, $x_0=0.3$, $N=20$, and $p_l=0.4$. With the help of Theorem \ref{them1}, the optimal targeted advertising strategy and corresponding high-quality access service market share can be obtained, shown in Fig. \ref{fig1}. If the MSP does not use advertising, the service market share will decrease quickly. Moreover, the advertising response constant is critical to the MSP's profits. The $x(t)$ increases with the increase of time $t$ when ${\eta _1}=2$, but decreases when ${\eta _1}=0.4$. The reason is that larger ${\eta _1}$ means the same targeted advertising investment can obtain greater effect. Furthermore, the more budget an MSP has for targeted advertising, the faster the acceleration bandwidth can be sold for the high-quality access service. Note that $T_2$ decreases when the basic bandwidth requirement to meet the user's minimal Meta-Immersion increases.

Figure \ref{fig2} shows that the user's downlink CR and noise versus the allocated bandwidth, with $m_{s_{ak}}=2$, $m_{ak}=3$, ${{\mu _{jk}}}=2$, $ {{\alpha _{jk}}}=2$, $p_a=10$ ${\rm dBW}$, $ {{{\bar \gamma }_{jk}}}=5$ ${\rm dBW}$, and different $ {{{\bar \kappa }_{ak}}} $. The warden adjusts its detection threshold, $ \varepsilon  $, dynamically and the jammer adjusts the jamming power to ensure the covertness of the communication, which means that we always have $ \left( 1-{\xi _w}\right) < \delta$, where $\delta$ is an arbitrarily small number. Here we set $ \delta=3\%$. We can observe that, when $ {{{\bar \kappa }_{ak}}} $ is larger, which means that the channel condition is better, we can allocate less bandwidth to the user to achieve the minimum requirement of downlink covert rate. As shown in Fig. \ref{fig2}, the increase in bandwidth also brings an increase in noise, which should be considered in some noise sensitive scenes.

Figure \ref{fig3} depicts the uplink BER versus the transmit power of HMD, with $ {{{\bar \kappa }_{ka}}}=10$ ${\rm dBW}$, $ {\sigma _{ka}^2}=1$ ${\rm dBW}$, different channel parameters and modulations. We can observe that the error rate decrease $80.7\%$ when the transmit power increases from $6$ ${\rm dBW}$ to $8$ ${\rm dBW}$, when DPSK is used, ${\it m_{ka}}=4$, and ${\it m_{s_{ka}}}=4$. Moreover, the channel condition has a great impact on the uplink performance. From Fig. \ref{fig3}, it can be observed that the minimal required transmit power decreases $9$ ${\rm dBW}$ when ${\it m_{s_{ka}}}$ and ${\it m_{ka}}$ increase $1$ and $2$, respectively. Changing the modulation format from BFSK to DPSK can decrease the required transmit power by $1.2$ ${\rm dBW}$. Considering that the users' HMDs are typically powered by batteries, reducing HMDs' power consumption indicates less heat generation, which can enhance users' Metaverse experiences.
\section{Conclusion}
We proposed a bandwidth allocation and targeted advertising scheme for wireless edge MSP, under budget constraints. After allocating basic bandwidth to users for normal-quality access, the MSP offers high-quality access services to make more profits. With the help of targeted advertising, acceleration bandwidth sales were boosted and the existence of advertisements cannot be detected by competitors. To determine the basic bandwidth, we proposed a novel metric, Meta-Immersion, to represent the user's feelings in the Metaverse. Several technical indicators, such as the DEP, downlink CR, and uplink BER under different modulation schemes were derived. We observed that the channel conditions of users can highly affect the experience in the Metaverse, and targeted advertising plays an important role in the sale of acceleration bandwidth. For future directions, many issues about advertising and Metaverse access can be studied, such as joint optimization of resource pricing and advertising.
	\bibliographystyle{IEEEtran}
	\bibliography{IEEEabrv,Ref}
	
\end{document}